\documentclass[conference]{IEEEtran}
\ifCLASSINFOpdf
\else
\fi
\usepackage{tikz}
\usepackage[sumlimits,intlimits]{amsmath}
\usepackage{amsmath,amsfonts,amssymb}%
\usepackage{bm}%
\usepackage{graphicx,graphics}%
\usepackage{cases}%
\usepackage[noadjust]{cite}%
\usepackage{color}%

\usepackage{verbatim}
\usepackage{enumerate}
\usepackage{algorithm}
\usepackage{balance}
\usepackage{multirow}
\usepackage{stfloats}
\usepackage{blindtext}
\usepackage{xspace}
\usepackage{booktabs}
\usepackage{subfig}
\usepackage{cite}
\usepackage[justification=centering]{caption}
\usepackage[belowskip=-7pt,aboveskip=0pt]{caption}
\begin{document}

\title{ A Compact and High Gain Dielectric-Loaded
	60GHz Multi-Stepped Waveguide Antenna Array}

\author{\IEEEauthorblockN{Saeideh Shad,  Hani Mehrpouyan}
	\IEEEauthorblockA{Department of Electrical and Computer Engineering\
		, Boise State Univeristy\\
		Email: saeidehshad@boisestate.edu, hanimehrpouyan@boisestate.edu }
	
}


%


\maketitle

\vspace{-4cm}
\begin{abstract}
	
	In this paper, a wideband high-gain $2\times 2$-element subarray is presented for 60 GHz applications. The antenna is fed with waveguide-fed cavity backed configuration and designed entirely via simple rectangular apertures. To improve radiation pattern characteristics and reduce the antenna size, stepped-radiating apertures loaded with a solid dielectric material. A standard WR-15 rectangular waveguide is designed to excite the antenna at the input port over the operation frequency.  The most significant advantage of using this design is its efficient radiation patterns, ability to decrease complexity and cost of fabrication. Simulated results demonstrate a maximum gain of about 19.5 dB, and the sidelobe level (SLL) of the antenna is less than -19 dB in E- and H-planes radiation patterns over the frequency range from 57.5 to 62.5 GHz. In contrast to previous works, the proposed antenna is much more simple to use in antenna array applications. Reduction in the number of radiating apertures and compact feeding networks will lower significantly the size and complexity of a large array with higher gain.
\end{abstract}

 \begin{keywords}
 	Rectangular waveguide, aperture antenna, high-gain antennas, millimeter-wave.
 \end{keywords}


%
\IEEEpeerreviewmaketitle
\vspace*{0cm}
\section{Introduction}

 Design of high-gain millimeter-wave (mm-wave) antennas is crucial for realizing mm-wave wireless communication systems. The strong oxygen absorption and the high propagation path loss in this frequency band  is affected the development of mm-wave antennas \cite{citation1,citation2}. Thus, design requirements for such antennas include highly directional pattern with high radiation efficiency.  Antennas with high gain produce very directive narrow beam that can overcome severe path loss at mm-wave frequencies. Based on this demand, different kinds of high gain broadband antennas such as microstrip arrays, substrate-integrated waveguide (SIW) arrays, and corporate-fed waveguide slot array antennas have been a subject of extensive research  \cite{citation3,citation4,citation5}. This letter presents a new profile waveguide aperture antenna with high gain, high efficiency and wide impedance bandwidth for the 60 GHz band applications. The antenna consists of a multilayer structure that are electrically connected together. The simulation results show that the proposed antenna covers  impedance bandwidth of 5 GHz from 57.5 to 62.5 GHz. As a result, this antenna could be used as the basic element of a large array antenna.

\section{Design and Configuration}

\begin{figure}[!]
\centering
	\includegraphics[width=80mm]{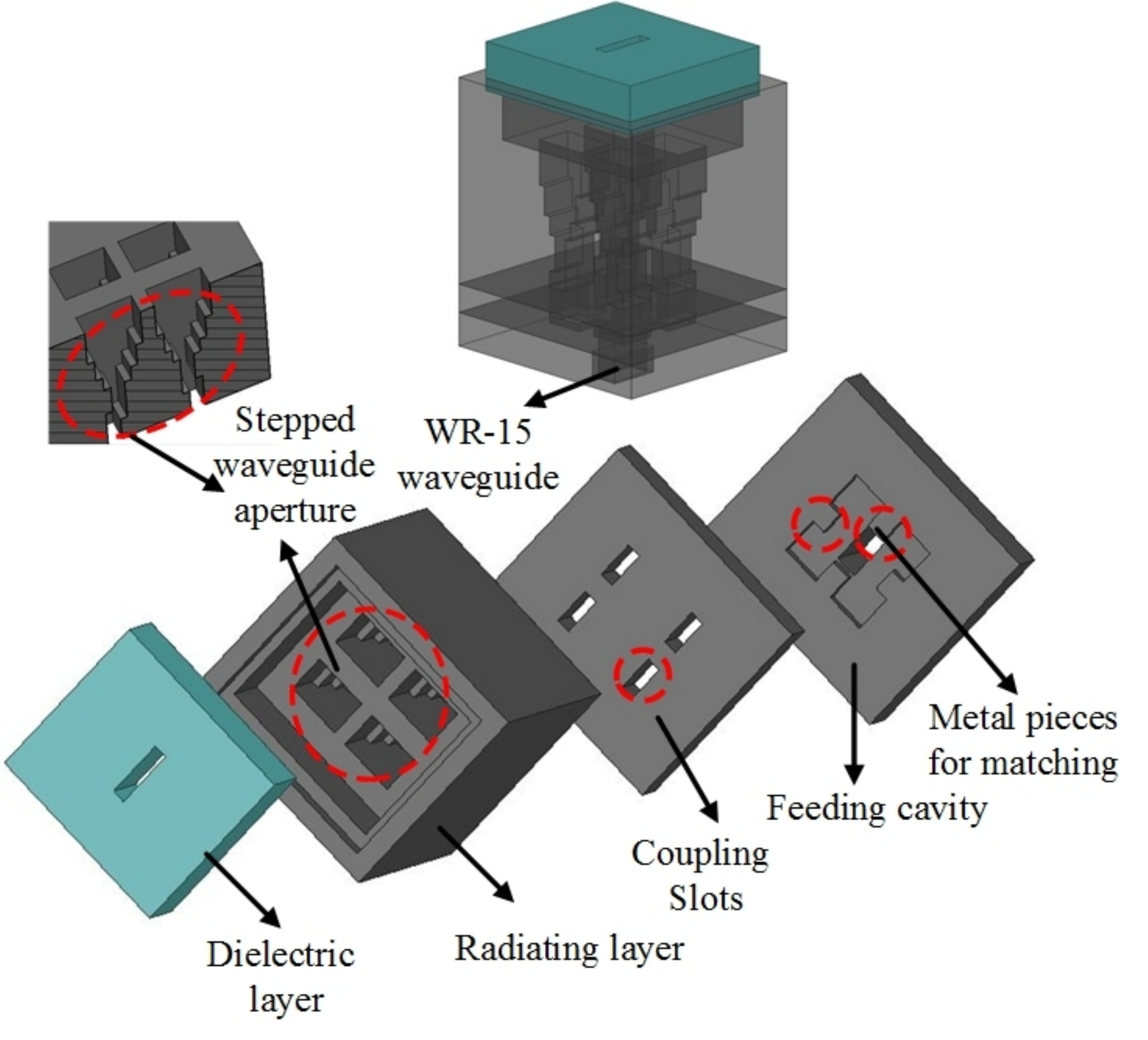}
	\caption{3-D view of the proposed antenna.}
	\label{fig:view}
	
\end{figure}

Fig. \ref{fig:view} shows the configuration of the proposed multilayer $2\times 2$-element subarray antenna. The antenna consists of three layers, i.e., feeding-cavity,  radiating layer, and dielectric layer. The radiating layer is composed of a set of four stepped rectangular waveguide designed to be symmetrical with respect to the center of the feeding-cavity. The aperture of the antenna is loaded with a dielectric material proposed to improve cross-polarization, aperture efficiency and gain characteristics without increasing the size of the antenna. The overall antenna is fed via a standard WR-15 rectangular waveguide at the bottom surface of the antenna to couple energy to the feeding cavity. As shown in Fig. \ref{fig:view}, the coupling is performed through the coupling slots at the top of the feeding cavity mainly used to couple power to the radiating layer.  In this case, the coupling slot in the feeding cavity have the same size as a standard WR-15 rectangular waveguide. Metallic pieces in the feeding cavity are utilized to adjust the matching between the cavity and  radiating layer.

\begin{table*}[!]
	\renewcommand{\arraystretch}{1.2}
	
	\caption{Radiation characteristics of the antenna}
	\label{tab2}
	\centering
	\begin{tabular} {cccccc}

		\hline\hline
		
		& E-plane & H-plane & E-plane & H-plane &  \\ [-0.5mm]  
		Frequency & 3 dB Beamwidth & 3 dB Beamwidth & First Sidelobe Level & First Sidelobe Level & Gain\\ [-0.5mm] 
		[GHz] & [deg.] & [deg.] & [dB] & [dB] & [dB] \\
		
		\hline
		
		58 & 21.5 & 18.95 & -21.3 & -27& 19.25 \\
		60 & 20.3 & 18.7 & -23.7 & -19.3 &19.4 \\
		62 & 18.9 & 18.6 & -26 & -19 & 19.16 \\
		\hline	
		\vspace{-30pt}
	\end{tabular} 
	
\end{table*}
  
\begin{figure}[ht]
	\centering
	
	\includegraphics[width=87mm]{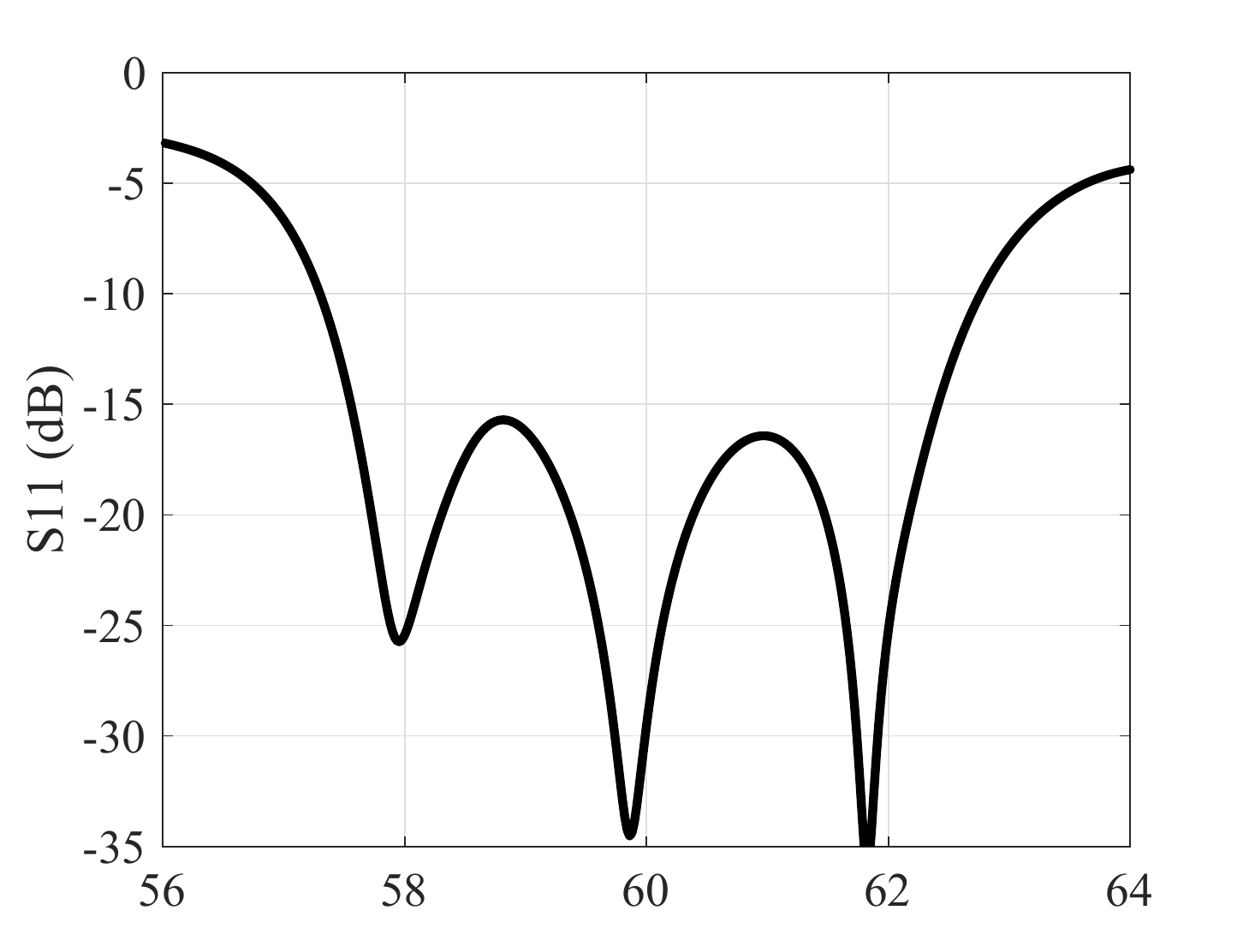}
	\caption{Simulated Reflection coefficient of $2\times 2$ subarray antenna.}
	\label{fig:S11}
		
\end{figure} 

\begin{figure}[] 
	
	\centering
	\includegraphics[width=90mm]{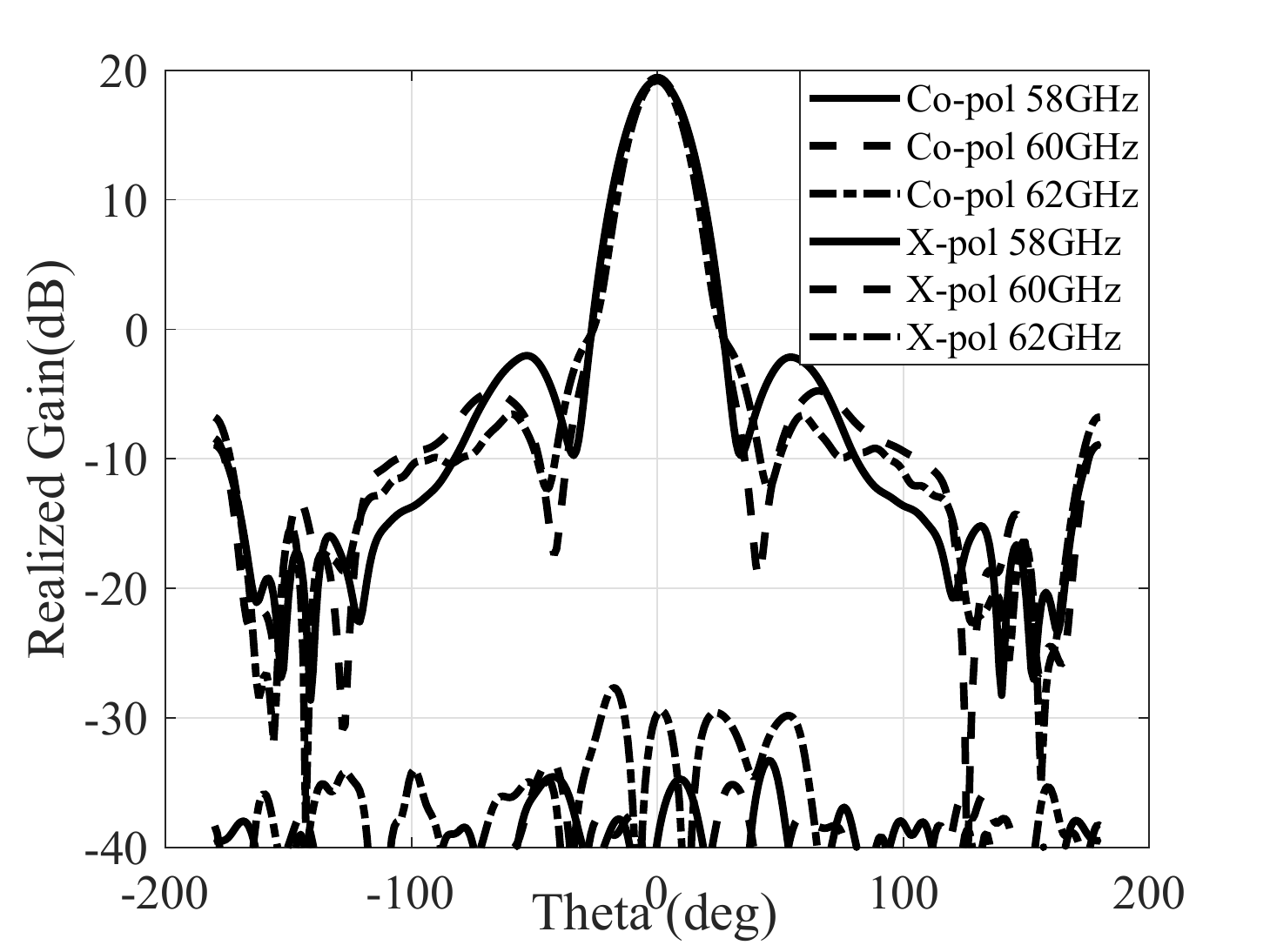}
	\caption{Radiation patterns of the simulated $2\times 2$ subarray antenna at E-plane and frequencies 58, 60, and 62 GHz.}

	\label{fig:phi0}
	
\end{figure}

\section{Smulation Results}

The designed subarray is analyzed using full-wave simulation software Ansoft HFSS. A parametric study of the impedance and radiation characteristics of the antenna has been performed over a frequency range of 57-64 GHz. The designed variables of the layers were carefully optimized to maintain a broadband impedance matching with high-gain and high-efficiency performance. This has been accomplished by optimizing the dimension of stepped waveguide apertures, metallic walls inside the feeding cavity, and dielectric material. Ultimately, the antenna is designed with the size of $20\times18\times19$ $mm^3$. Fig.\ref{fig:S11} depicts the simulated reflection coefficient of the antenna. It can be seen that the simulated {-}10 dB bandwidths of the antenna is 9\% from 57.3 to 62.7 GHz. The radiation pattern of the antenna in the E- and H-planes at three different frequencies 58, 60, 62GHz are shown in Figs. 3 and 4. The half-power beamwidth, first sidelobe level, and antenna gain discrimination are shown in Table II as a function of the frequency. Moreover, the cross-polarization values are less than -27 dB in the two planes.  The frequency behavior of the simulated gain is shown in Fig. \ref{fig:gain}. It can be observed that the realized gain varies from 18.6 to 19.5 dB over the frequency band of 57.5 to 62.5 GHz.

\section{Conclusion}
\vspace{0cm}
 we have designed and studied a new structure of the waveguide antenna array for 60 GHz applications. In addition, we propose a new design where the radiating apertures of the waveguide is loaded with a dielectric material to increase antenna gain and aperture efficiency. Reflection coefficient, gain, aperture efficiency, sidelobe and cross-polarization characteristics are studied. The simulated realized gain is higher than 18.5 dB over the entire operation bandwidth from 57.5 to 62.5 GHz, corresponding to sidelobe level less than -19dB.

\begin{figure}[]
	\centering

	\includegraphics[width=90mm]{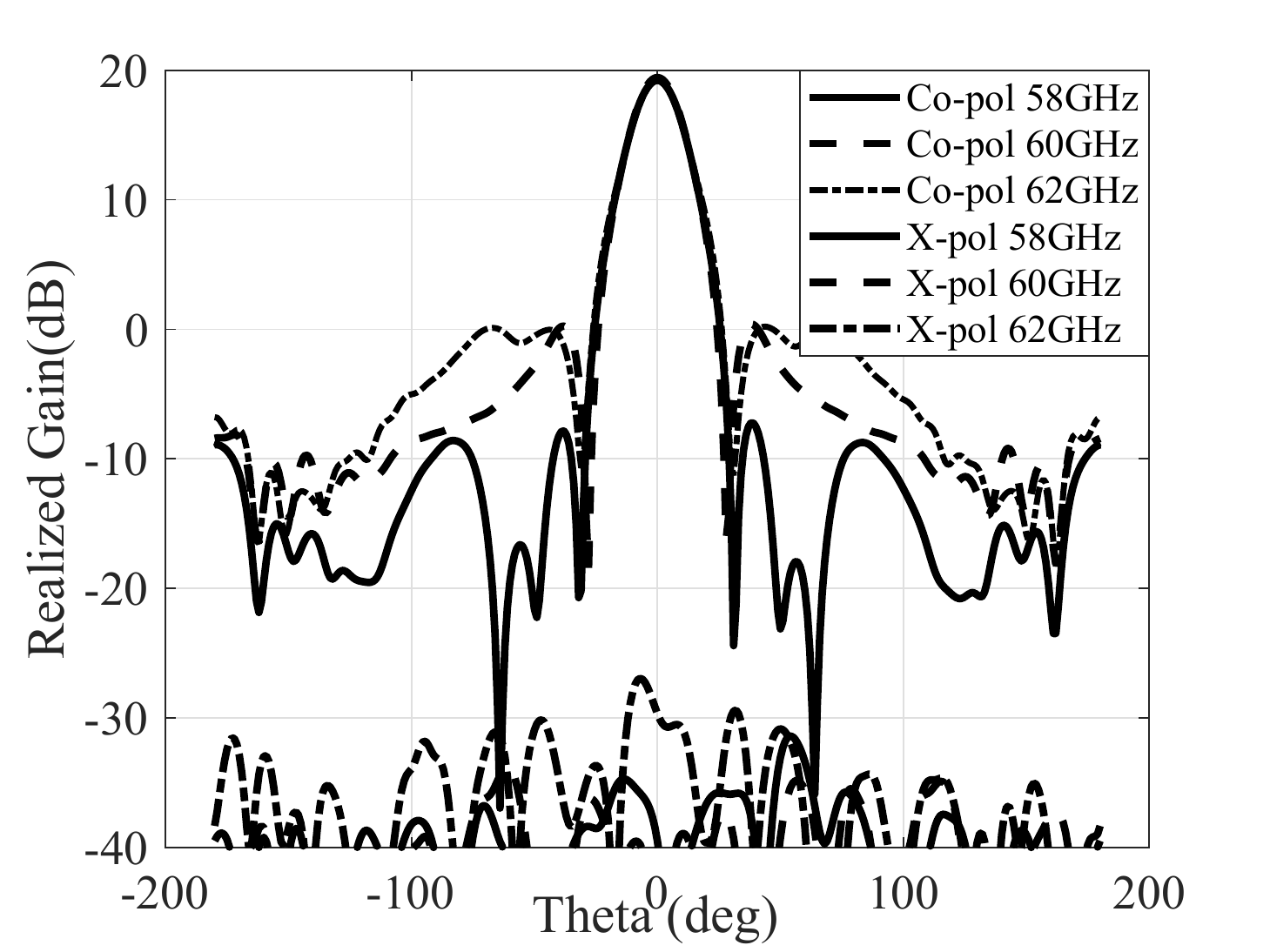}
	\caption{Radiation patterns of the simulated $2\times 2$ subarray antenna at H-plane and frequencies 58, 60, and 62 GHz.}
	\label{fig:phi90}
\end{figure}

\begin{figure} []
	\centering

	\includegraphics[width=90mm]{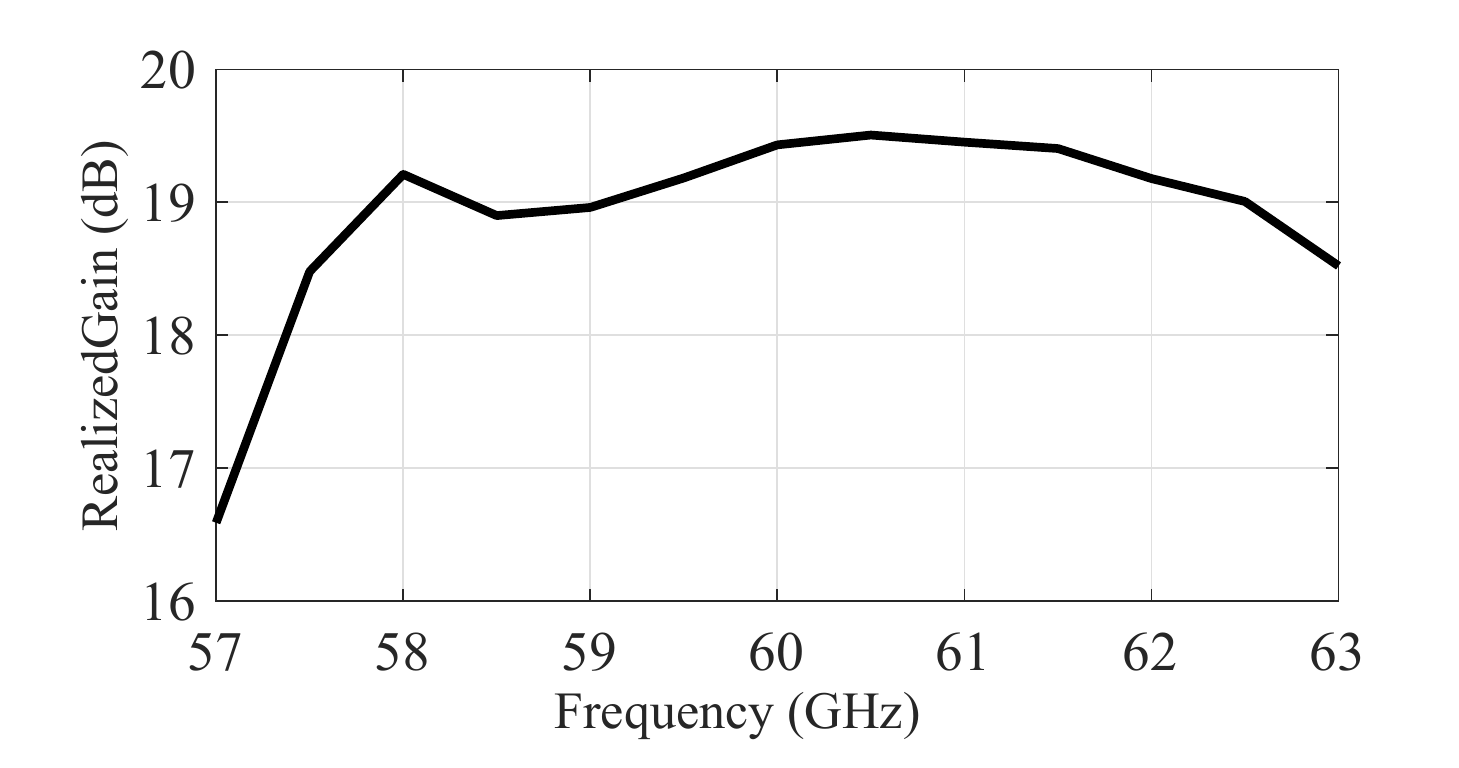}
	\caption{Simulated gain of $2\times 2$ subarray antenna.}
	\label{fig:gain}
\end{figure}





\begin{thebibliography}{11}
\bibitem{citation1}
 Y. Wang et al., “5G Mobile: Spectrum Broadening to Higher-Frequency
Bands to Support High Data Rates,” IEEE
Vehic. Tech. Mag., vol. 9, no. 3, Sept 2014, pp. 39–46.

\bibitem{citation2}
R. Amiri and H. Mehrpouyan, “Self-organizing mm-wave networks: A
power allocation scheme based on machine learning,” in Proc. IEEE
GSMM, pp. 1–4, May 2018.
\bibitem{citation3}
M. Gholami et al, “A Compact and High-Gain Cavity-Backed Waveguide Aperture Antenna in the C-Band for High-Power Applications,” IEEE Trans. Antennas Propag., vol. 66, no. 3, pp.
1208-1216, Mar 2018.

\bibitem{citation4}
 D. Zarifi et al, “Design and
fabrication of a high-gain 60-GHz corrugated slot antenna array with
ridge gap waveguide distribution layer,” IEEE Trans. Antennas Propag.,
vol. 64, no. 7, pp. 2905–2913, Jul. 2016.

\bibitem{citation5}
W. El-Halwagy, R. Mirzavand, J. Melzer, M. Hossain, and P. Mousavi,
‘‘Investigation of wideband substrate-integrated vertically-polarized electric dipole antenna and arrays for mm-wave 5G mobile devices,’’ IEEE
Access, vol. 6, pp. 2415–2457, 2018


\end{thebibliography}
%

\end{document}